\documentclass[conference]{IEEEtran}
\IEEEoverridecommandlockouts
\usepackage{cite}
\usepackage{amsmath,amssymb,amsfonts}
\usepackage{algorithmic}
\usepackage{graphicx}
\usepackage{textcomp}
\usepackage{xcolor}
\usepackage{url}
\def\BibTeX{{\rm B\kern-.05em{\sc i\kern-.025em b}\kern-.08em
    T\kern-.1667em\lower.7ex\hbox{E}\kern-.125emX}}

\begin{document}

\title{Quantum-Train Long Short-Term Memory: Application on Flood Prediction Problem
}

\author{
\IEEEauthorblockN{
    Chu-Hsuan Abraham Lin\IEEEauthorrefmark{3}\IEEEauthorrefmark{7}, Chen-Yu Liu \IEEEauthorrefmark{1}\IEEEauthorrefmark{2}\IEEEauthorrefmark{6}, 
    Kuan-Cheng Chen\IEEEauthorrefmark{4}\IEEEauthorrefmark{5}\IEEEauthorrefmark{8}
}

\IEEEauthorblockA{\IEEEauthorrefmark{1} Hon Hai (Foxconn) Research Institute, Taipei, Taiwan}
\IEEEauthorblockA{\IEEEauthorrefmark{2}Graduate Institute of Applied Physics, National Taiwan University, Taipei, Taiwan}
\IEEEauthorblockA{\IEEEauthorrefmark{3}Department of Electrical and Electronic Engineering, Imperial College London, London, UK}
\IEEEauthorblockA{\IEEEauthorrefmark{4}Department of Materials, Imperial College London, London, UK}
\IEEEauthorblockA{\IEEEauthorrefmark{5}Centre for Quantum Engineering, Science and Technology (QuEST), Imperial College London, London, UK}

\IEEEauthorblockA{Email:\IEEEauthorrefmark{6} chen-yu.liu@foxconn.com, \IEEEauthorrefmark{7} abraham.lin23@imperial.ac.uk, \IEEEauthorrefmark{8}kuan-cheng.chen17@imperial.ac.uk}

}

\maketitle

\begin{abstract}
Flood prediction is a critical challenge in the context of climate change, with significant implications for ecosystem preservation, human safety, and infrastructure protection. In this study, we tackle this problem by applying the Quantum-Train (QT) technique to a forecasting Long Short-Term Memory (LSTM) model trained by Quantum Machine Learning (QML) with significant parameter reduction. The QT technique, originally successful in the “A Matter of Taste” challenge at QHack 2024, leverages QML to reduce the number of trainable parameters to a polylogarithmic function of the number of parameters in a classical neural network (NN). This innovative framework maps classical NN weights to a Hilbert space, altering quantum state probability distributions to adjust NN parameters. Our approach directly processes classical data without the need for quantum embedding and operates independently of quantum computing resources post-training, making it highly practical and accessible for real-world flood prediction applications. This model aims to improve the efficiency of flood forecasts, ultimately contributing to better disaster preparedness and response.

\end{abstract}


\section{Introduction}

In recent years, the convergence of climate science and technology has spurred innovations to tackle natural disasters, such as river flooding, which pose significant risks to ecosystems, human life, and infrastructure. The Deloitte Quantum Climate Challenge 2024 integrates Quantum Machine Learning (QML) with flood forecasting to enhance prediction models using quantum technologies, underscoring the critical importance of this issue. 
Quantum neural networks (QNNs) in Quantum Machine Learning (QML) utilize superposition and entanglement to process multiple outcomes simultaneously, potentially speeding up training and promising advancements across various fields \cite{qml1, qml2, qml3, qml4, qmlapp1, qmlapp2, qgan1, qmlapp3, qgan2, qrs1, qgan3, qgan4, rlqls, qnnlearn1, qnnlearn2, liu2022hybrid, qnnlearn3, learnqpe, qnnlearn4, qnnlearn5, qnntrain1, qnntrain2, rsnn, liu2023practical}. However, despite its potential, QML faces significant challenges. Practical application issues, such as scaling problems with gate angle encoding, arise because larger input sizes require wider and deeper quantum circuits, which can reduce accuracy in the Noisy Intermediate-Scale Quantum (NISQ) era. Classical preprocessing can mitigate these issues by reducing data dimensions but may lose vital information, necessitating improved data encoding methods. Additionally, the requirement for quantum computing hardware during the inference stage limits widespread application, as quantum computing resources are limited and expensive.

The Long Short-Term Memory (LSTM) model \cite{lstm1} in classical machine learning (ML) is one of the most promising approaches to tackle time-series forecasting problems. Consequently, an active area of research is exploring how a quantum version of LSTM \cite{qlstm} could enhance LSTM, similar to the comparison between ML and QML. As previously mentioned, directly replacing classical layers with quantum layers faces challenges related to data encoding and the requirement for quantum hardware during inference. Overcoming these challenges could significantly enhance the power of QML applications. This is where the Quantum-Train (QT) approach \cite{tcnnqc2, qt1, qtrl} comes into play. The QT framework “trains” the classical neural network (NN) by mapping classical NN weights to a Hilbert space and adjusting quantum neural network (QNN) parameters to modify classical NN parameters. This model not only reduces trainable parameters to a polylogarithmic scale but also processes classical data directly, operating independently of quantum resources post-training, thus enhancing practicality and accessibility.

\section{Method}
To address the challenge of flood prediction based on historical observation data, we can frame this task as a time series prediction problem. The LSTM model is chose due to its proficiency in handling long sequences and its ability to remember information over extended periods. After providing an overview of the LSTM model, we will train it using the QT technique, with much fewer training parameters. The trained classical model will then be employed to predict flood events, leveraging its enhanced capabilities from the quantum-assisted training process. This approach aims to improve the accuracy and reliability of flood predictions.

\begin{figure}[htp]
\centering
\includegraphics[scale = 0.18]{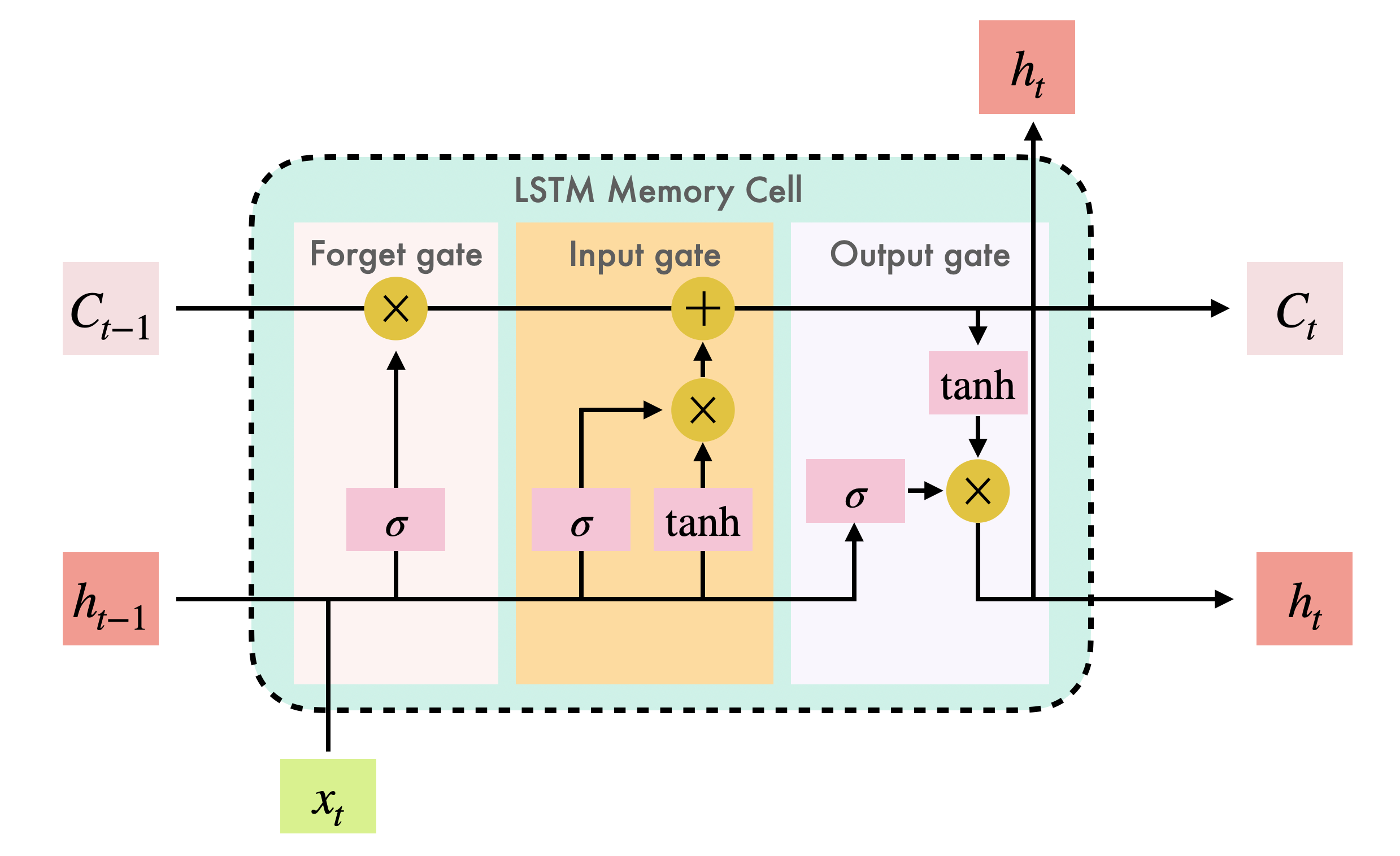}
\caption{
A typical illustration of LSTM cell.} 
\label{fig:lstm}
\end{figure}

\subsection{LSTM}
LSTM model is a specialized form of Recurrent Neural Network (RNN) originally introduced by Sepp Hochreiter and Jürgen Schmidhuber in 1997 \cite{lstm1}. It was designed to address the vanishing gradient problem that affects standard RNNs during backpropagation, particularly over long input sequences. Unlike traditional RNNs that struggle to maintain information across many timesteps, LSTMs can carry information across long sequences, making them ideal for tasks involving sequential data such as language processing, time series analysis, and speech recognition. An LSTM unit typically comprises three gates: the input gate, the forget gate, and the output gate, which regulate the flow of information. These gates control the extent to which new data enters the cell, old data is forgotten, and current data is used to compute the output activation of the LSTM unit. This structure allows the LSTM to effectively learn from data with long-range temporal dependencies, overcoming the limitations of earlier RNN designs. A typical LSTM visualization is illustrated in Fig.~\ref{fig:lstm}.


\subsection{Quantum-Train LSTM}
In this section, we outline the key elements of the Quantum-Train LSTM (QT-LSTM) framework, clarify the relationship between classical NNs and quantum states, and discuss the parameterization and training of these quantum states.
\subsubsection{Mapping Quantum State to Classical LSTM}
We begin with a classical NN, specifically an LSTM, defined by the parameter vector $\vec{\theta}$:
\begin{equation}
\vec{\theta} = (\theta_1, \theta_2, \ldots, \theta_M),
\end{equation}
where $M$ is the number of parameters. We introduce a quantum state $|\psi\rangle$ encoded by $N = \lceil \log_2 M \rceil$ qubits, sufficient to cover $M$ parameters. The probabilities of measuring the quantum state, represented by $|\langle i | \psi \rangle|^2$, range between 0 and 1 for $i \in \{1, 2, \ldots, 2^N\}$. The main goal is to link these quantum measurement probabilities to the parameters of the classical NN. Specifically, we extract the parameter $\theta_i$ from the quantum probabilities $|\langle i | \psi \rangle|^2$.

A mapping model $M_{\vec{\gamma}}$, based on an additional NN with parameter vector $\vec{\gamma}$, is used. Inputs to this model include binary-encoded basis information combined with measured probabilities. For example, an input vector $\vec{x}_i$ for a 7-qubit scenario could be:
\begin{equation}
\vec{x}_i = [1,0,1,1,1,0,0,0.058],
\end{equation}
where $|\langle 10111000| \psi \rangle|^2 = 0.058$. The mapping model $M_{\vec{\gamma}}$ maps this input to a corresponding parameter $\theta_i$ in the classical NN:
\begin{equation}
M_{\vec{\gamma}}(\vec{x}_i) = \theta_i.
\end{equation}
The length of the input to the mapping model is $N+1$, and the dimension of $\vec{\gamma}$ varies within $\text{poly}(N)$, depending on the polynomial number of layers in the mapping NN.
\subsubsection{Quantum Neural Networks}
\begin{figure*}[htp]
\centering
\includegraphics[scale = 0.22]{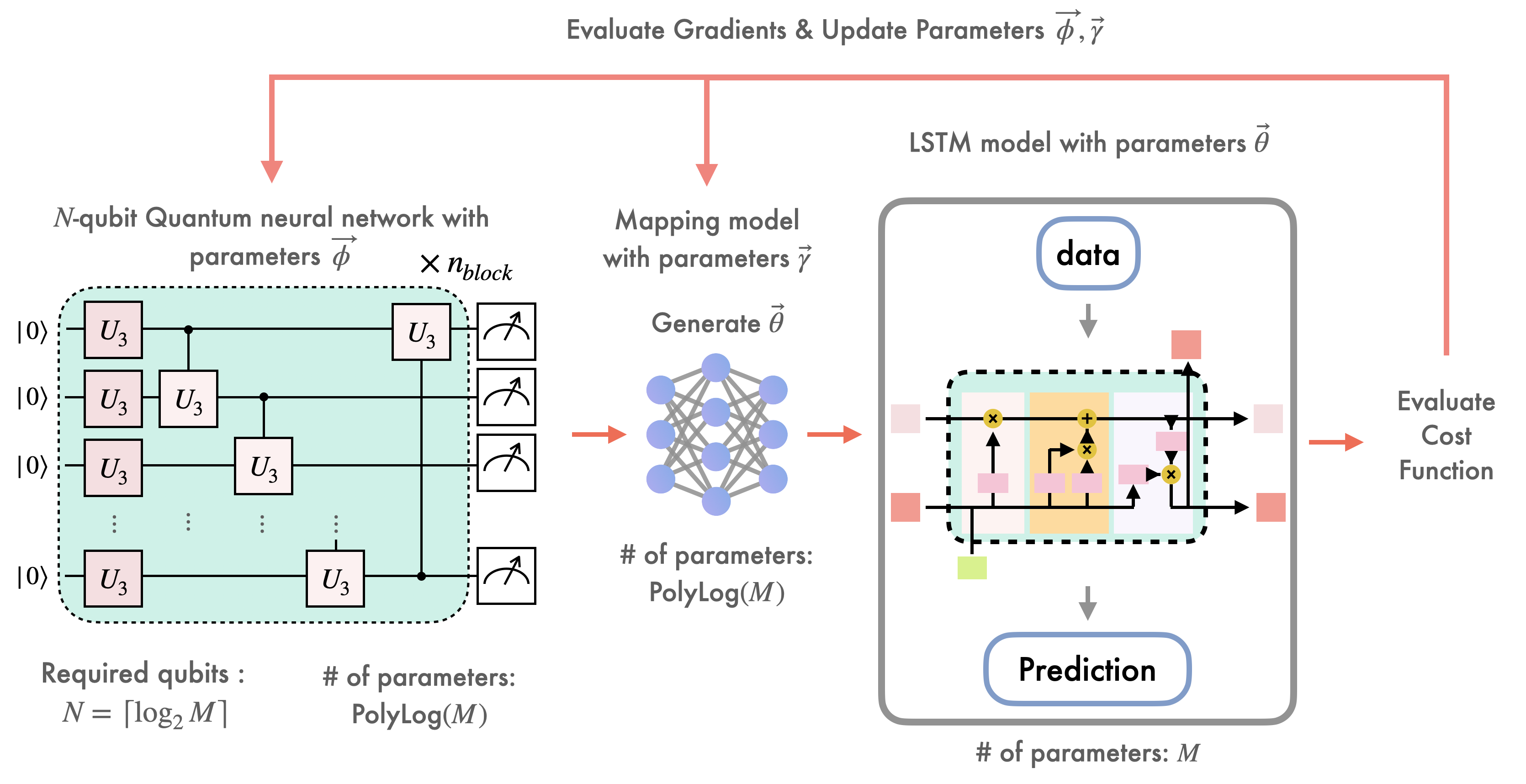}
\caption{Quantum-Train LSTM.}
\label{fig:scheme}
\end{figure*}

We now explore the construction of the quantum state $|\psi \rangle$, which is parameterized by rotational gate angles and expressed as $|\psi(\vec{\phi}) \rangle$. This parameterized quantum state forms the QNN \cite{pqcml1}, developed based on a specific ansatz defining the quantum circuit configuration. We use the $U_3$ gate, essential for precise quantum state adjustments:
\begin{equation}
U_3(\mu, \varphi, \lambda) = \left[ \begin{array}{cc}
\cos(\mu/2) & -e^{i \lambda} \sin(\mu/2) \\
e^{i \varphi} \sin(\mu/2) & e^{i(\varphi + \lambda)} \cos(\mu/2)
\end{array} \right]
\end{equation}
The controlled-$U_3$ gate ($CU_3$) is used for entangling qubits:
\begin{equation}
CU_3 = I \otimes |0\rangle \langle 0 | + U_3(\mu, \varphi, \lambda) \otimes |1\rangle \langle 1 |
\end{equation}
The number of parameters for this QNN scales polynomially with the qubit count. Fig.~\ref{fig:scheme} visually represents our QNN ansatz, ensuring the parameters scale as $O(\text{polylog}(M))$.

\subsubsection{Training Flow for Quantum-Train LSTM}
The QT-LSTM model starts with configuring an $N$-qubit QNN with parameterized $U_3$ gates. These blocks can be repeated $n_{\text{block}}$ times to enhance the model’s ability to map quantum states to classical parameters. QNN parameters $\vec{\phi}$ derive classical LSTM weights via the mapping model with parameters $\vec{\gamma}$.

The classical LSTM network, defined by parameter $\vec{\theta}$, undergoes conventional evaluaion, involving data ingestion, processing, and output prediction. The quality of predictions is evaluated using the Mean-Square-Error (MSE) loss function:
\begin{equation}
\label{eq:mseloss}
\ell_{MSE} = \frac{1}{N_{\text{d}}} \sum_{n=1}^{N_{\text{d}}} (y_n - \hat{y}_n)^2,
\end{equation}
where $y_n$ represents the actual data and $\hat{y}n$ the predictions, with $N{\text{d}}$ being the total number of training samples. The goal is to minimize this function through optimization, computing gradients for $\vec{\phi}$ and $\vec{\gamma}$. Gradients are calculated analytically in numerical simulations. These gradients update $\vec{\phi}$ and $\vec{\gamma}$, refining the NN parameters for improved performance.

As illustrated in Fig.~\ref{fig:scheme}, the QT-LSTM framework is efficient, requiring fewer QNN parameters that scale as $O(\text{polylog}(M))$. This method enables the trained model to be used on classical computers for inference tasks, significantly enhancing practicality, especially when quantum computing resources are limited.

\section{Result and Discussion}


Building on the QT-LSTM method outlined in the previous section, this section presents the results of our flood prediction analysis. We utilized a comprehensive set of features in our input data, including water levels and discharge rates from five river stations along the Wupper, as well as volume and fill levels from four water reservoirs. Additionally, we incorporated various weather data and forecasts from three weather stations.

To enhance our model's predictive accuracy, we introduced four supplementary columns representing different time lags in the Kluserbrücke water levels, ranging from 1 to 7 data points. These time lags help the LSTM model capture temporal tendencies which are crucial to an accurate forecast.

The LSTM model inputs consist of 30 data points of these features, spanning several hours to two days—depending on the frequency of measurements—to predict the maximum water level for the following 24 hours.

\subsection{Water Level Forecasting}
\begin{figure*}[htp]
\centering
\includegraphics[scale = 0.25]{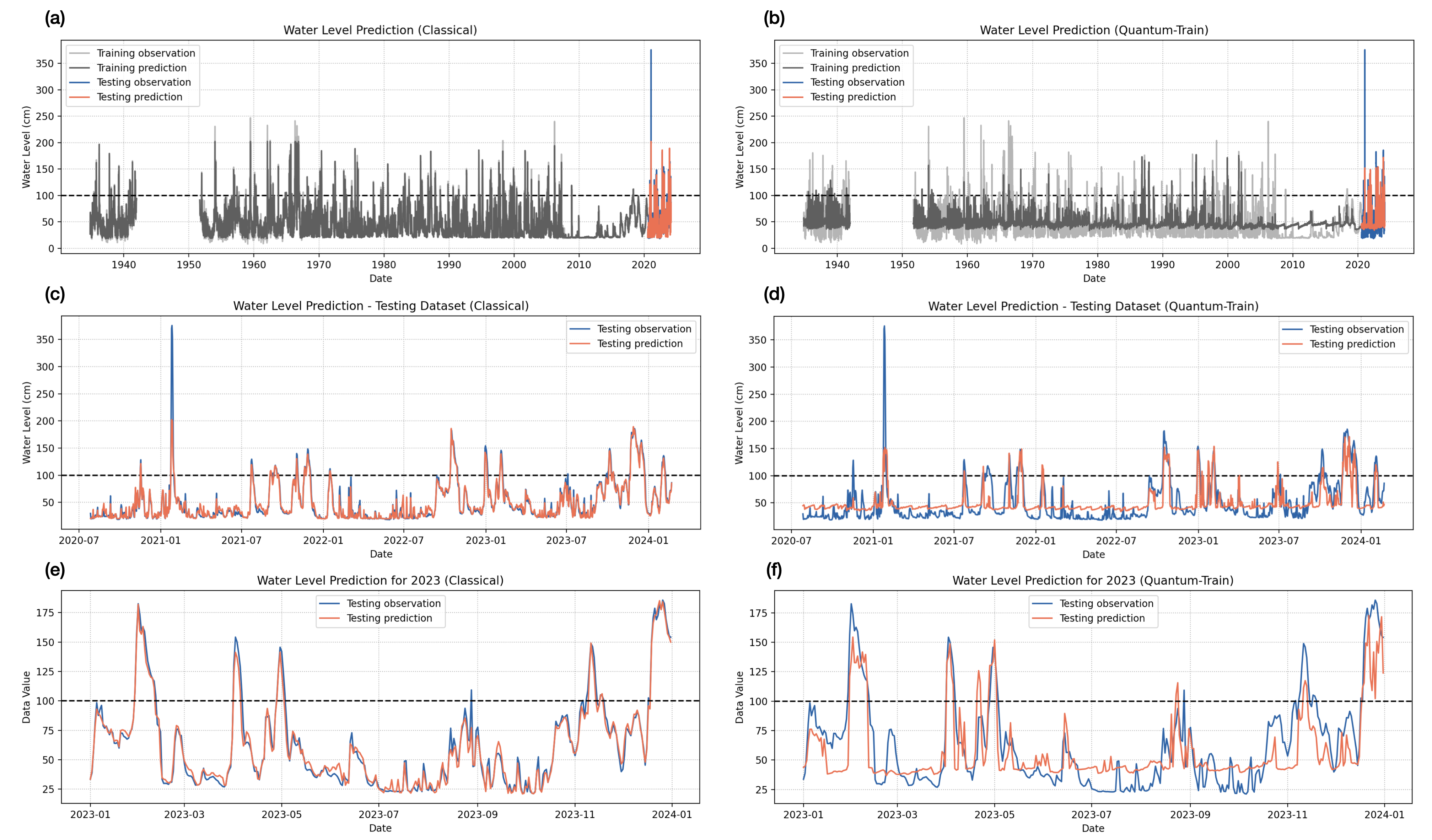}
\caption{
Comparative analysis of water level predictions between classical and QT methods. (a) and (b) provide the performance of models over the whole dataset, where the QT method exhibits lower prediction accuracy despite a reduced parameter requirement. (c) and (d) zoom into the testing dataset, and (e) and (f) detail predictions for the year 2023, highlighting the trade-off between the QT method's parameter efficiency and its predictive performance.} 
\label{fig:res_1}
\end{figure*}

Forecasting water levels accurately is paramount for effective flood management and mitigating the impacts of climate change. Employing an LSTM architecture, the classical method has established its efficacy in water level forecasting with a Mean Squared Error (MSE) of 38.02 and Mean Absolute Error (MAE) of 4.19, using a total of 40,451 parameters. as shown in Fig.~\ref{fig:res_2}(a). 

The QT-LSTM requries $\lceil \log_2 {40451} \rceil = 16$ qubits in this example. The QT-LSTM, as shown in Fig.~\ref{fig:res_2}(a), exhibits a higher MSE of 475.91 and MAE of 17.44 in its performance. The outcome of the classical model and QT-LSTM is presented in Fig.~\ref{fig:res_1}(a). With the polylog reduction of training parameters to train the same classical LSTM model, the QT-LSTM approach requires significantly fewer parameters (18,830), below half of that of the classical model (40,451), which theoretically suggests a more streamlined model. For the training time, it is worth noting that the resulting 11 hours is from the simulation of the quantum computation on the classical computer, and could not reflect the real situation of the training time when using a real quantum computer in the future. 

In comparison, Fig.~\ref{fig:res_1}(b), (d), (e) display the QT-LSTM method's predictions over the same period as Fig.~\ref{fig:res_1}(a), (c), (e). Despite the less accurate predictions during the testing phase, the visual similarity in patterns suggests that the difference in accuracy may not be drastic.

\subsection{Classification Prediction}
\begin{figure*}[htp]
\centering
\includegraphics[scale = 0.25]{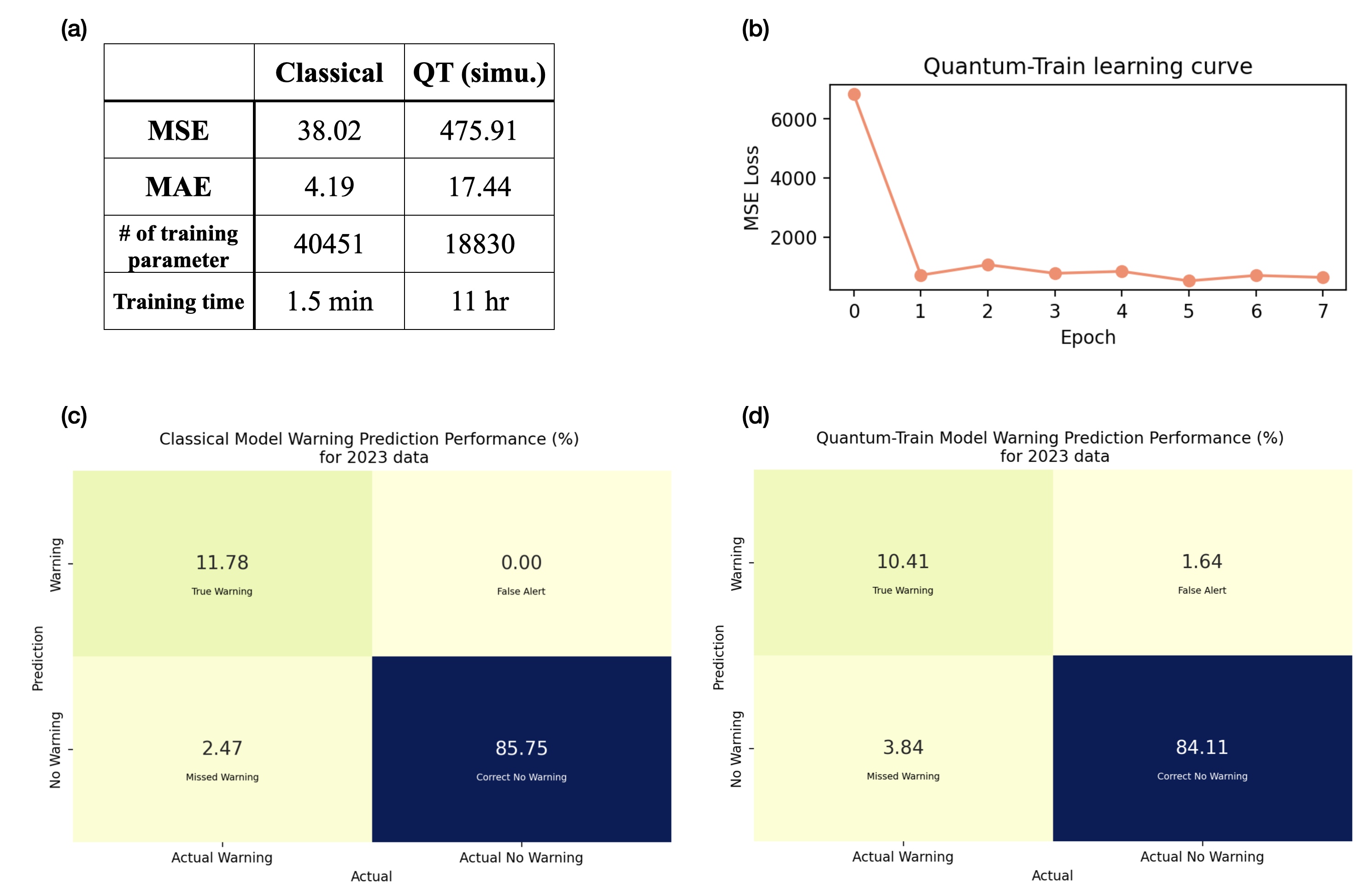}
\caption{
(a) Performance of classical LSTM and QT-LSTM. (b) Learning curve for QT-LSTM. (c) and (d): Comparison of flood warning prediction performance for 2023 data between classical and QT models, highlighting the percentage of true warnings, false alerts, missed warnings, and correct no-warning predictions.} 
\label{fig:res_2}
\end{figure*}

Based on the water level forecasting result, with a threshold of 100 cm for flood events, we can extract the occurrence of flood events and process the classification predictions. In Fig.~\ref{fig:res_2}(c) and (d), the performance of the classical and QT-LSTM models in predicting flood warnings for the year 2023 is provided. The performance is measured in terms of percentage and showcases the models' abilities to correctly predict actual warnings and avoid false alerts.

For the classical model, we observe that it correctly issues a warning (True Warning) 11.78\% of the time and correctly predicts the absence of a warning (Correct No Warning) 85.75\% of the time. This model did not generate any false alerts (0.00\% False Alert) and missed warnings 2.47\% of the time (Missed Warning).

In comparison, the QT-LSTM model issued accurate warnings slightly less frequently at 10.41\% (True Warning) and had a marginally lower performance in correctly predicting the absence of a warning at 84.11\% (Correct No Warning). This model produced a small percentage of false alerts, at 1.64\% (False Alert), and missed warnings 3.84\% of the time (Missed Warning).

The outcomes indicate that both the classical and QT-LSTM models demonstrate a proficient ability to predict flood occurrences and non-occurrences. However, the classical model marginally surpasses the QT-LSTM model in accuracy, evidenced by a lower rate of missed warnings and an absence of false alerts. It is particularly noteworthy that despite the classical method's superior MSE and MAE in water level prediction, the practical application of forecasting flood events reveals a negligible difference in effectiveness between the two models.  Notably, the QT-LSTM model achieves a comparable level of overall performance while utilizing significantly fewer training parameters.

\subsection{Scalability and Practicality}
In earlier research, we applied the QT techniques for the task of classification \cite{tcnnqc2}, exploring their efficacy within the realm of QML. Following this, we have now extended our investigation to LSTM models, and our initial findings are promising. These results demonstrate that integrating QT with LSTM models is viable, enhancing the adaptability and application spectrum of QT methods in QML. The versatility of QT is further evidenced by its application across different modeling paradigms from classification tasks to more complex temporal dynamics of LSTM models, which are essential for tasks like flood forecasting. Despite being slightly less accurate in water level prediction compared to classical methods, QT performs reasonably in the critical use case of flood event prediction. The ability of QT to maintain comparable performance with classical methods, despite a significant reduction in the number of training parameters, highlights its potential for efficiency and scalability. These attributes are particularly advantageous in scenarios where computational resources are limited or where the quantum infrastructure is nascent yet growing.

As we continue to refine and expand the capabilities of QT, it is becoming increasingly clear that this approach holds significant promise for the field of QML. By adapting QT to work in concert with LSTM architectures, we are paving the way for more resource-efficient and broadly applicable ML models that are capable of tackling the diverse challenges presented by big data and complex predictive tasks.

In pure QML, data encoding into quantum states poses a challenge due to the constraints on input data size. More complex data necessitates additional gate angles for encoding, leading to potentially deep quantum circuits for larger datasets. An alternative approach involves preprocessing and compressing the data but at risk of losing vital information. In contrast, the QT methodology leverages classical inputs and outputs, effectively mapping quantum states to classical NN parameters. This approach retains the computational benefits of Hilbert space while circumventing the drawbacks associated with data encoding in quantum circuits.

Furthermore, both pure QML and hybrid QML models necessitate a quantum computer for executing QNN during the inference stage. In contrast, the QT architecture outcome is a trained purely classical model that can operate independently on classical hardware, bridging the quantum hardware gap associated with traditional QML, and offering a practical solution for a wide application.

\section{Summary}
In this research, we employ QT method to train classical. By leveraging the computational principles of quantum mechanics, this approach aims to reduce the training parameters for classical LSTM models, which are crucial in forecasting flood events—an increasingly important task in the face of climate change. 

We detail a novel methodology where classical LSTMs, known for their efficacy in processing temporal sequences, are trained through a quantum-assisted framework. This hybrid model significantly reduces the requisite number of trainable parameters by utilizing the compact nature of quantum state representations. The core innovation of our approach is the mapping of probabilities associated with quantum states to the parameters of classical NNs, allowing for an efficient hybrid quantum-classical training process. Empirical results demonstrate the QT-LSTM model's ability to predict water levels, a critical factor in flood forecasting. Although the QT model shows a slight decrease in accuracy compared to classical methods, the substantial reduction in parameter count and the independence from quantum computing resources during inference significantly boost its practicality and scalability. Further discussions highlight the broader implications of the QT approach, particularly its capacity to efficiently manage large-scale models with limited quantum resources. This is vital in the context of the ongoing expansion of model sizes across the machine learning landscape, which poses significant computational challenges.

QT-LSTM framework not only maintains the essential predictive performance of classical models in some cases but also offers enhanced efficiency and adaptability. This makes it a viable and promising solution for addressing complex ML problems, especially in scenarios constrained by computational resources.

\section{Acknowledgments} 
This research is based on the winning project by HHRI TeamQC in the Deloitte Quantum Climate Challenge 2024.

\bibliographystyle{IEEEtran}
\bibliography{references}

\end{document}